\documentclass[12pt]{iopart}

\usepackage{graphicx}
\usepackage{fancyhdr}
\usepackage{amsfonts,amssymb,color}

\begin{document}

\title[]{Anomaly-free vector perturbations with holonomy corrections 
in loop quantum cosmology}

\author{Jakub Mielczarek$^1$, Thomas Cailleteau$^2$, Aurelien Barrau$^2$ and Julien Grain$^3$}

\address{$^1$ Astronomical Observatory, Jagiellonian University, 30-244
Krak\'ow, Orla 171, Poland}
\address{$^2$ Laboratoire de Physique Subatomique et de Cosmologie, UJF, CNRS/IN2P3, INPG\\
53, av. des Martyrs, 38026 Grenoble cedex, France}
\address{$^3$ Institut d'Astrophysique Spatiale, Universit\'e Paris-Sud 11, CNRS \\ B\^atiments 120-121, 91405 Orsay Cedex, France}

\begin{abstract}
We investigate vector perturbations with holonomy corrections in the framework 
of loop quantum cosmology. Conditions to achieve anomaly freedom for these 
perturbations are found at all orders. This requires the introduction of counter-terms
in the hamiltonian constraint. We also show that anomaly freedom requires the 
diffeomorphism constraint to hold its classical form when scalar matter is added
although the issue of a vector matter source, required for full consistency, remains 
to be investigated.The gauge-invariant variable and the corresponding equation of 
motion are derived. The propagation of vector modes through the bounce is finally 
discussed. 
\end{abstract}

\maketitle

\section{Introduction}

In the canonical formulation of general relativity, the Hamiltonian
is a sum of constraints. In particular, within the Asthekar
framework \cite{Ashtekar:1986yd}, the Hamiltonian is a sum of three constraints: 
\begin{equation}
H_{\rm G}[N^i,N^a,N] = \frac{1}{2\kappa } \int_{\Sigma}d^3x \left( N^i C_i+N^a C_a+N C\right) \approx 0,
\nonumber 
\end{equation}
where $\kappa = 8\pi G$, $(N^i,N^a,N)$ are Lagrange multipliers,  $C_i$ is 
called the Gauss constraint,  $C_a$ is the diffeomorphism  constraint, and $C$ is 
the hamiltonian constraint.  The sign "$\approx$" means equality on the surface of 
constraints ({\it i.e.} weak equality). One can also define the corresponding smeared 
constraints as follows: 
\begin{eqnarray}
\mathcal{C}_1 &=& G[N^i] =\frac{1}{2\kappa } \int_{\Sigma}d^3x\ N^i C_i, \\
\mathcal{C}_2 &=& D[N^a] =\frac{1}{2\kappa } \int_{\Sigma}d^3x\ N^a C_a, \\
\mathcal{C}_3 &=& S[N] =\frac{1}{2\kappa } \int_{\Sigma}d^3x\ N C,
\end{eqnarray}
that is such that $H_{\rm G}[N^i,N^a,N] = G[N^i] +D[N^a]+S[N]$. The Hamiltonian is a 
total constraint which is vanishing for all multiplier functions $(N^i,N^a,N)$. 

Because $H_{\rm{G}}[N^i,N^a,N] \approx 0$ at all times, the time derivative
of the Hamiltonian constraint is also weakly vanishing, $\dot{H}_{\rm G}[N^i,N^a,N] \approx 0$. 
The Hamilton equation $\dot{f}=\{f,H_{\rm G }[M^i,M^a,M]\}$ therefore leads to
\begin{equation}
\left\{H_{\rm G }[N^i,N^a,N], H_{\rm G }[M^i,M^a,M]\right\} \approx 0, \label{HH}
\end{equation}
which, when explicitly written, means:
\begin{equation}
\left\{ G[N^i] +D[N^a]+S[N], G[M^i] +D[M^a]+S[M] \right\} \approx 0. \nonumber
\end{equation}
Due to the linearity of the Poisson bracket, one can straightforwardly find that the condition 
(\ref{HH}) is fulfilled if the smeared constraints belong to a first class algebra
\begin{equation}
\{ \mathcal{C}_I, \mathcal{C}_J \} = {f^K}_{IJ}(A^j_b,E^a_i) \mathcal{C}_K. \label{algebra}
\end{equation}
In (\ref{algebra}), the ${f^K}_{IJ}( A^j_b,E^a_i)$ are structure functions which, in general, 
depend on the phase space (Ashtekar) variables  $(A^j_b, E^a_i)$. The algebra 
of constraints is fulfilled at the classical level due to general covariance. 
To prevent the system from escaping the surface of constraints, leading to an
unphysical behavior, the algebra must also be closed at the quantum level.  
In addition, it was pointed out in \cite{Nicolai:2005mc} 
that the algebra of quantum constraints should be strongly closed (\emph{off shell} closure). 
This means that the relation (\ref{algebra}) should hold in
the whole kinematical phase space, and not only on the surface of constraints 
(\emph{on shell} closure). This should remain true after promoting the 
constraints to quantum operators.  

Loop quantum gravity (LQG) \cite{Ashtekar:2004eh} is a promising approach to 
quantize gravity, based on a canonical formalism parametrized by  Ashtekar variables.  
The methods of LQG applied to cosmological models are known as loop quantum 
cosmology (LQC) \cite{Bojowald:2008zzb}. In LQC,  quantum gravity effects are introduced 
by holonomies of Ashtekar connection. This replacement is necessary because 
connection operators do not exist in LQG. Rewriting classical constraints in terms 
of holonomies leads to two types of quantum corrections: the so-called inverse-volume 
and holonomy corrections. Because the constraints are quantum-modified, the corresponding 
Poisson algebra might not be closed:  
\begin{equation}
\{ \mathcal{C}^Q_I, \mathcal{C}^Q_J \} = {f^K}_{IJ}(A^j_b,E^a_i) \mathcal{C}^Q_K+
\mathcal{A}_{IJ}. 
\end{equation}
Here, $\mathcal{A}_{IJ}$ stands for the anomaly term which can appear due to the 
quantum modifications. For consistency (closure of algebra), $\mathcal{A}_{IJ}$ is 
required to vanish. The condition $\mathcal{A}_{IJ}=0$ implies some restrictions on 
the form of the quantum corrections. In this paper, we will study this requirement to find a 
consistent way for introducing quantum holonomy corrections to the vector perturbations.
 
The question of the construction of an anomaly-free algebra of constraints is especially interesting to 
address in inhomogeneous LQC. Perturbations around 
the cosmological background are indeed responsible for structure formation in the Universe. 
This  gives a chance to link quantum gravity effects with  astronomical
observations. In the particular case of the flat FLRW background, the Ashtekar variables
can be decomposed as follows
\begin{equation}
 A^i_a = \gamma \bar{k} \delta^i_a +\delta A^i_a \ \ \ {\rm and} \ \ \ E_i^a = \bar{p} \delta_i^a +\delta E_i^a, 
\end{equation}
where $\bar{k}$ and $\bar{p}$ parametrize the background phase space, and 
$\gamma$ is the so-called Barbero-Immirzi parameter. 

The issue of anomaly freedom for the algebra of cosmological perturbations was extensively 
studied for  inverse-volume corrections. It was shown that this requirement
can be fulfilled for first order perturbations. 
This was derived for scalar \cite{Bojowald:2008gz,Bojowald:2008jv}, 
vector \cite{Bojowald:2007hv} and tensor perturbations \cite{Bojowald:2007cd}. 
It is worth mentioning that, for the tensor perturbations, the anomaly-freedom is 
automatically satisfied.  Based on the anomaly-free scalar perturbations, predictions 
for the power spectrum of cosmological perturbations were also performed 
\cite{Bojowald:2010me}. This gave a chance to put constraints on 
some parameters of the model using observations of the cosmic microwave 
background radiation (CMB) \cite{Bojowald:2011hd}.  
  
The aim of this article is to address the issue of anomaly freedom for the holonomy-corrected 
vector perturbations in LQC. It was shown in \cite{Bojowald:2007hv} that 
these perturbations can be anomaly free up to the fourth order in the canonical variable 
$\bar{k}$. This, however, is not sufficient to perform the analysis of the propagation of vector modes 
through the cosmic bounce. Vector perturbations with \emph{higher order 
holonomy corrections} were also recently studied \cite{Li:2011zzd}. It was shown there that, 
in this case, an anomaly-free formulation can be found for the gravitational sector. 
In this paper, we apply a different method, which is based on the introduction of 
counter-terms in the Hamiltonian constraint. We show that the  anomaly freedom conditions for 
vector modes with holonomy corrections can be fulfilled in this way. The method is similar 
to the one used by Bojowald \emph{et al.} in the case of cosmological perturbations 
with  inverse-volume  corrections. As we will see, the counter-terms do not introduce any 
higher-order holonomy corrections. This way of fulfilling the anomaly freedom 
is therefore different from what was done in \cite{Li:2011zzd}, where higher order terms 
are 
involved. Moreover,  in Ref.  \cite{Li:2011zzd}, the issue of anomaly freedom was 
studied for the gravity sector only and the formulation suffers from ambiguities. In our
study, scalar matter is introduced. The presence of this matter term fixes the
ambiguity associated with the holonomy correction.
It should be underlined that without a vector matter source, one cannot rigorously 
prove the anomaly cancellation. However, as will be shown in the next sections, 
our approach is meaningful as the equations derived are, as in \cite{Bojowald:2007hv},
compatible with vector matter assuming $\pi_a=0$ but $V_a\neq 0$.

Holonomy corrections arise while regularizing classical constraints, when expressing  the 
Ashtekar connection in terms of holonomies. In particular, the regularization of the curvature of the 
Ashtekar connection $F_{ab}^i$ leads to the factor 
$\left( \frac{\sin(\bar{\mu} \gamma \bar{k})}{\bar{\mu}\gamma} \right)^2$, which simplifies to 
$\bar{k}^2$ in the classical limit $\bar{\mu} \rightarrow 0$. However, the Ashtekar connection
does not appear only because of $F_{ab}^i$: in the classical perturbed constraints, terms linear 
in $\bar{k}$ are also involved. In principle, such terms should be holonomy-corrected.
However, there is no direct expression for them, analogous to the regularization
of the  $F_{ab}^i$ factor. Nevertheless, one can naturally expect that $\bar{k}$ factors are corrected 
by the replacement \footnote{this was derived rigorously e.g. for the 
Bianchi II model \cite{Ashtekar:2009um}.}  
\begin{equation}
\bar{k} \rightarrow \frac{\sin(n\bar{\mu} \gamma \bar{k})}{n\bar{\mu}\gamma},
\end{equation}
where $n$ is some unknown integer. It should be an integer because, when quantizing the theory, 
the $e^{i\gamma \bar{k}}$ factor is promoted to be the shift operator 
acting on the lattice states.  If $n$ was not an integer, the action of the  operator corresponding  
to $e^{i n\gamma \bar{k}}$ would be defined in a different basis. Another issue 
is related with the choice of  $\bar{\mu}$, which corresponds to the so-called 
\emph{lattice refinement}. Models with a power-law parametrization
$\bar{\mu} \propto \bar{p}^{\beta}$ were discussed in details in the literature. While, in general, 
$ \beta \in [-1/2,0]$, it was pointed out that the choice $\beta = -1/2$ is favored 
\cite{Nelson:2007um}. This particular choice is called the $\bar{\mu}-$scheme 
(new quantization scheme). Studies in this article are performed for the 
general power-law case $\bar{\mu} \propto \bar{p}^{\beta}$.

For the sake of simplicity,  we use the notation  
\begin{equation}
\mathbb{K}[n] := \left\{  
\begin{tabular}{ccc} 
$\frac{\sin(n\bar{\mu} \gamma \bar{k})}{n\bar{\mu}\gamma}$  &  for &$n \in \mathbb{Z}/\{0\}$,  \\ 
& & \\
 $\bar{k}$  &  for & $n=0$,
\end{tabular}
\right.
\end{equation}
for the holonomy correction function. The introduction of holonomy 
corrections is therefore performed by replacing $\bar{k}  \rightarrow \mathbb{K}[n]$. 
However, factors $\bar{k}^2$ are simply replaced by $\mathbb{K}[1]^2$, because
they arise from the curvature of the Ashtehar connection. For the linear terms, the 
integers are parameters to be fixed.  

\section{Vector perturbations with holonomy corrections}

Vector modes within the canonical formulation were studied in \cite{Bojowald:2007hv}. 
It was shown there that 
\begin{equation}
\delta E^a_i = - \bar{p} (c_1 \partial^aF_i+ c_2 \partial_iF^a), \label{deltaE}
\end{equation}
where $c_1+c_2=1$ and the divergence-free condition $\delta^i_a \delta E^a_i =0$ is fulfilled. 
The values of $c_1$ and $c_2$ depend on the gauge choice. However, due to the Gauss constraint, 
only symmetric variables are invariant under internal rotations. This is the case for 
$\delta {E^{(a}}_{i)} $, which is consequently independent on the specific choice of $c_1$ and $c_2$, 
and should be preferred. The perturbation of the shift vector is parametrized as $\delta N^a = S^a$.
  
We consider the quantum holonomy-corrected hamiltonian constraint  given by 
\begin{equation}
S^Q[N] = \frac{1}{2 \kappa} \int_{\Sigma} d^3x \left[ \bar{N}(C^{(0)}+C^{(2)})\right],  \label{scalar}
\end{equation}
where 
\begin{eqnarray}
C^{(0)} &=& -6\sqrt{\bar{p}}  \left(\mathbb{K}[1] \right)^2,  \\
C^{(2)} &=&  -\frac{1}{2\bar{p}^{3/2}}\left( \mathbb{K}[1]  \right)^2(1+\alpha_1) (\delta E^c_j \delta E^d_k \delta^k_c \delta^j_d) 
+  \sqrt{\bar{p}}(\delta K^j_c \delta K^k_d \delta^c_k \delta^d_j)  \nonumber \\
&-&\frac{2}{\sqrt{\bar{p}}}\left(\mathbb{K}[v_1]\right)(1+\alpha_2)(\delta E^c_j \delta K^j_c ). 
\end{eqnarray}
Holonomy corrections were introduced by replacing $\bar{k}  \rightarrow \mathbb{K}[n]$. Two counter-term
functions $\alpha_1$ and $\alpha_2$, whose interest will be made clear later, were also added. In the classical limit $\mathbb{K}[n]  \rightarrow \bar{k}$, 
and $\alpha_i=\alpha_i(\bar{p},\bar{k}) \rightarrow 0$, with $i=1,2$. We have assumed here that 
$\alpha_i$ are functions of the background variables only and that
$v_1$ is an integer to be fixed. The hamiltonian constraint (\ref{scalar}) corresponds to the one 
investigated in \cite{Bojowald:2007hv} while setting $\alpha_i=0$. However, as we will show, it is necessary 
to introduce these additional factors, which vanish in the classical limit. These factors can, of course, also be viewed as 
contributions from the two counter-terms    
\begin{eqnarray}
S_{C1} &=&  -\frac{\alpha_1}{2 \kappa} \int_{\Sigma} d^3x\frac{\bar{N} }{2\bar{p}^{3/2}}\left(\mathbb{K}[1]  \right)^2 (\delta E^c_j \delta E^d_k \delta^k_c \delta^j_d),  \\
S_{C2} &=& -\frac{\alpha_2}{2 \kappa} \int_{\Sigma} d^3x  \frac{2 \bar{N}}{\sqrt{\bar{p}}}\left(\mathbb{K}[v_1]\right)
(\delta E^c_j \delta K^j_c) 
\end{eqnarray}
to the holonomy-corrected hamiltonian constraint. 

A similar method of counter-terms was successfully applied for perturbations 
with inverse-volume corrections.  In that case, it was possible to fix the counter-terms so as to make the algebra anomaly free. In this article, we follow the same path so as to find 
explicit expressions for $\alpha_1$ and $\alpha_2$. 

For the sake of completeness, we also introduce holonomy corrections to the diffeomorphism constraint, as follows:  
\begin{eqnarray}
D^Q[N^a] = \frac{1}{\kappa} \int_{\Sigma}  d^3x \delta N^c  \left[ -\bar{p} (\partial_k \delta K^k_c )  
 - \left( \mathbb{K}[v_2]\right)\delta^k_c  (\partial_d \delta E^d_k ) \right],  \label{diff}
\end{eqnarray}
where $v_2$ is an unknown integer.  It is worth emphasizing here that within LQG, the 
diffeomorphism constraint is fulfilled at the classical level while constructing the diffeomorphism 
invariant spin network states. If LQC was really derived from the full LQG theory, the 
classical form of the diffeomorphism constraint should therefore be used. However, at this early
stage of the understanding of LQC, it might be safe to allow for some
generalizations by introducing the holonomy correction also to the diffeomorphism constraint.  
This hypothesis was already studied in \cite{Wu:2010wj} in the case of 
holonomy-corrected scalar perturbations. It was assumed there that the 
holonomy correction function was  given by $\mathbb{K}[2]$. In this work, we prefer to keep a more general 
expression $\mathbb{K}[v_2]$ with a free $v_2$ parameter. We will investigate whether this 
additional modification can help to fulfill the anomaly freedom conditions.

In order to investigate the algebra of constraints, the Poisson bracket has to be defined. 
We start with the gravity sector for which the Poisson bracket 
can be decomposed as follows: 
\begin{eqnarray}
\left\{ \cdot , \cdot \right\} &=& \frac{\kappa}{3V_0} \left( \frac{\partial \cdot }{\partial \bar{k}} \frac{\partial \cdot}{\partial \bar{p}}
- \frac{\partial \cdot}{\partial \bar{p}} \frac{\partial \cdot }{\partial \bar{k}}  \right)  \nonumber \\
&+& \kappa \int_{\Sigma} d^3x \left( \frac{\delta \cdot}{\delta \delta K^i_a }\frac{\delta \cdot}{\delta \delta E^a_i } 
-\frac{\delta \cdot}{\delta \delta E^a_i }\frac{\delta \cdot}{\delta \delta K^i_a } \right). 
\label{Poisson}
\end{eqnarray}

The algebra of constraints (\ref{scalar}) and (\ref{diff}) shall now be investigated. Using the Poisson 
bracket (\ref{Poisson}), we find:
\begin{eqnarray}
\left\{S^Q[N_1], S^Q[N_1] \right\} &=&  0, \\
\left\{D^Q[N^a_1], D^Q[N^a_2] \right\} &=& 0,  \\ 
\left\{S^Q[N], D^Q[N^a]\right\} &=& \frac{\bar{N}}{\sqrt{\bar{p}}} \mathcal{B} D^Q[N^a] \nonumber \\
&+&\frac{\bar{N}}{\kappa\sqrt{\bar{p}}} \int_{\Sigma}  d^3x\delta N^c \delta^k_c  
(\partial_d  \delta E^d_k) \delta E^d_k \mathcal{A}, 
\end{eqnarray}
where $\mathcal{B}:=  (1+\alpha_2)\mathbb{K}[v_1]
+  \mathbb{K}[v_2]- 2  \mathbb{K}[2]$, and $\mathcal{A}$ is the anomaly function which, for reasons
that shall be made clear later, is decomposed
in two parts $\mathcal{A} = \mathcal{A}_1+\mathcal{A}_2$, where 
\begin{eqnarray}
\mathcal{A}_1 &=& \mathcal{B}\mathbb{K}[v_2],  \\
\mathcal{A}_2 &=& 2\mathbb{K}[2] \bar{p} \frac{\partial \mathbb{K}[v_2]}{\partial \bar{p}} 
- \frac{1}{2} (\mathbb{K}[1])^2 \cos(v_2 \bar{\mu} \gamma \bar{k}) 
                        - 2\mathbb{K}[1] \bar{p} \frac{\partial \mathbb{K}[1]}{\partial \bar{p}}\cos(v_2 \bar{\mu} \gamma \bar{k}) \nonumber \\
                        &+& (1+\alpha_2)\mathbb{K}[v_1] \mathbb{K}[v_2] -\frac{1}{2} \mathbb{K}[1]^2(1+\alpha_1).   
\end{eqnarray}
This decomposition was made such that, in the classical limit ($\bar{\mu} \rightarrow 0$), 
both contributions to the anomaly vanish separately.  Using the relation
\begin{equation}
\bar{p} \frac{\partial \mathbb{K}[n]}{\partial \bar{p}} = \left( \bar{k} \cos(n\bar{\mu} \gamma \bar{k}) -\mathbb{K}[n]\right)\beta,
\end{equation}
the second contribution can be re-written as:
\begin{eqnarray}
\mathcal{A}_2  &=& -2\beta \mathbb{K}[2] \mathbb{K}[v_2] + (1+\alpha_2)\mathbb{K}[v_1]\mathbb{K}[v_2] 
                        + (2\beta -1/2) (\mathbb{K}[1])^2\cos(v_2 \bar{\mu} \gamma \bar{k}) \nonumber \\
                        &-& \frac{1}{2}(\mathbb{K}[1])^2(1+\alpha_1). \label{A2rewritten}
\end{eqnarray}
The full 
anomaly term is given by:
\begin{eqnarray}
\mathcal{A} &=& 2(1+\alpha_2) \mathbb{K}[v_1]\mathbb{K}[v_2]-\frac{1}{2}(\mathbb{K}[1])^2(1+\alpha_1)
-2(1+\beta) \mathbb{K}[2] \mathbb{K}[v_2]+\mathbb{K}[v_2]^2  \nonumber \\
&+&(2\beta -1/2) (\mathbb{K}[1])^2\cos(v_2 \bar{\mu} \gamma \bar{k}).
\label{totalanomaly}
\end{eqnarray} 
 
\section{Anomaly freedom in the gravity sector} 

\begin{figure}[ht!]
\centering
\includegraphics[width=6cm,angle=0]{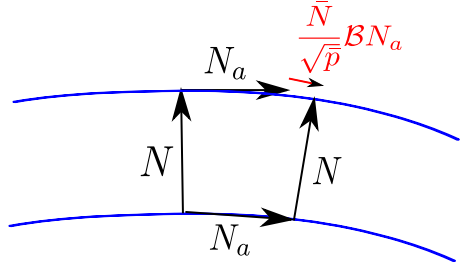}
\caption{Pictorial representation of the hypersurface deformation algebra.} 
\label{Fig}
\end{figure}

The requirement of the anomaly freedom for the gravity sector reads as $\mathcal{A} = 0$.
Under this condition, the algebra of constraints becomes closed but deformed, in particular: 
\begin{equation}
\left\{S^Q[N], D^Q[N^a]\right\} = D^Q\left[\frac{\bar{N}}{\sqrt{\bar{p}}} \mathcal{B}  N^a \right].
\end{equation}
The structure of space-time is therefore also modified. This is illustrate in Fig. \ref{Fig} where
one can notice that the hamiltonian and diffeomorphism constraints generate gauge transformations 
in directions respectively normal and parallel to the hypersurface.  
In the classical limit, $\mathcal{B} \rightarrow 0$ and both the transformations commute at the 
perturbative level. 
   
\subsection{The no counter-terms case}

Let us start by analyzing the condition $\mathcal{A} = 0$ without any counter-term 
({\it i.e.} with $\alpha_1=\alpha_2=0$). This case corresponds to the one studied in 
\cite{Bojowald:2007hv} generalized  by the contribution from the corrected 
diffeomorphism constraint. It was shown in that work that, if $v_2=0$, the 
anomaly-freedom condition can be satisfied up to the $\bar{k}^4$ order only. Here, we investigate whether
this might be improved by the additional correction made to the  diffeomorphism constraint.

By setting $\alpha_1=\alpha_2=0$, the anomaly term given by (\ref{totalanomaly}) 
can be expanded in powers of the canonical variable $\bar{k}$ as follows:
\begin{eqnarray}
\frac{\mathcal{A}}{(\bar{\mu} \gamma)^2} &=& \frac{1}{12} \left(20-4 v_1^2-v_2^2+8 \beta -8 v_2^2 \beta \right) x^4    \nonumber \\
 &+&\frac{1}{720} \left(-224+12 v_1^4-220 v_2^2+40 v_1^2 v_2^2 +17 v_2^4 \right. \nonumber \\
 &-& \left. 128 \beta +80 v_2^2 \beta +48 v_2^4 \beta \right) x^6 +\mathcal{O}(x^8),
\end{eqnarray}
where we have defined $x:=\bar{\mu} \gamma \bar{k}$ and  $x \in[0,\pi]$.
Clearly, in the classical limit $\bar{\mu}\rightarrow 0$, the anomaly  tends to zero. 
Requiring the anomaly cancellation up to the fourth order leads to the condition:
\begin{equation}
20-4 v_1^2-v_2^2+8 \beta-8 v_2^2 \beta = 0.
\end{equation}
It can be shown that the condition of anomaly cancellation up to orders higher 
than four cannot be met.  For $\beta=-1/2$ ($\bar{\mu}-$scheme), 
the  above equation simplifies to the quadratic Diophantine equation: 
\begin{equation}
16-4 v_1^2+3v_2^2 = 0.  \label{Diophantine}
\end{equation}
This equation can be reduced to a Pell-type equation and solved for an infinite number of  
pairs  of integers $(v_1,v_2)$. The first three solutions are $(2,0), (4,4)$ and $(14,16)$. 
The first one $(2,0)$ corresponds to the case studied in \cite{Bojowald:2007hv}, 
where the diffeomorphism constraint was kept at its classical form. The value $v_1=2$ 
obtained in this case was also used to fix the ambiguity for the holonomy-corrected tensor 
perturbations \cite{Bojowald:2007cd}. If the holonomy modified diffeomorphism 
constraint is used, the ambiguity cannot be fixed anymore due to the infinite number of solutions to 
Eq. (\ref{Diophantine}). 

As  we have shown, the modification of the diffeomorphism constraint does not help 
satisfying the anomaly freedom conditions in the absence of counter-terms. In this 
case, the anomaly freedom can be fulfilled up to the fourth order in $x$. In the 
semi-classical limit $x \ll 1$, the anomaly cancellation up to the fourth order might 
be a good approximation. 
However, when approaching the bounce, where $x=\frac{\pi}{2}$, contributions 
from higher order terms become significant and the effects of the anomaly cannot be 
neglected anymore. Studies of vector perturbations during the bounce phase cannot be 
performed in such a setup. In order to study vector perturbations through the bounce, 
the anomaly cancellation at all orders is required. This probably makes
mandatory the introduction of counter-terms.

\subsection{The general case}

Let us consider the general case with non-vanishing counter-terms. In this case, 
the requirement $\mathcal{A}=0$ can be translated into a relation between 
the two counter-terms 
$\alpha_1$ and $\alpha_2$:
\begin{eqnarray}
\alpha_1 &=&-1+4(1+\alpha_2) \frac{\mathbb{K}[v_1]\mathbb{K}[v_2]}{\mathbb{K}[1]^2} 
-4(1+\beta) \frac{\mathbb{K}[2] \mathbb{K}[v_2]}{\mathbb{K}[1]^2}+
2\frac{\mathbb{K}[v_2]^2}{\mathbb{K}[1]^2}  \nonumber \\
&+&(4\beta -1)\cos(v_2 \bar{\mu} \gamma \bar{k}).
\end{eqnarray} 
With this choice for the $\alpha_1$ function, the anomaly is removed. However 
a significant ambiguity remains. Namely, the function $\alpha_2$ together with 
parameters $v_1$ and $v_2$ remain undetermined. A particularly interesting case
corresponds to the choice $\alpha_2=0$. This determines $\alpha_1$. 
Of course, this also works the other way round: one can set $\alpha_1=0$ and  
derive the correct expression for $\alpha_2$.  Therefore, two special cases, heuristically motivated, where one of the 
counter-terms is vanishing, are worth studying:
\begin{eqnarray}
\alpha_1 &=&-1+4\frac{\mathbb{K}[v_1]\mathbb{K}[v_2]}{\mathbb{K}[1]^2} 
-4(1+\beta) \frac{\mathbb{K}[2] \mathbb{K}[v_2]}{\mathbb{K}[1]^2}+
2\frac{\mathbb{K}[v_2]^2}{\mathbb{K}[1]^2}  \nonumber \\
&+&(4\beta -1)\cos(v_2 \bar{\mu} \gamma \bar{k}), \\
\alpha_2 &=& 0,
\end{eqnarray}
and
\begin{eqnarray}
\alpha_1 &=& 0, \\
\alpha_2 &=&-1+ \frac{1}{4} \frac{(\mathbb{K}[1])^2}{\mathbb{K}[v_1]\mathbb{K}[v_2]} 
+(1+\beta)  \frac{\mathbb{K}[2]}{\mathbb{K}[v_1]}-\frac{1}{2}  \frac{\mathbb{K}[v_2]}{\mathbb{K}[v_1]}  \nonumber \\
&-&(\beta -1/4) \frac{(\mathbb{K}[1])^2\cos(v_2 \bar{\mu} \gamma \bar{k})}{\mathbb{K}[v_1]\mathbb{K}[v_2]}.
\end{eqnarray}
To conclude, at least one counter-term is necessary to fulfill the anomaly 
freedom conditions for the gravity sector.

\subsection{The $\mathcal{B}=0$ case}

Another possible way to fix the ambiguity in the choice of the
$\alpha_1$ and $\alpha_2$ functions could be to set $\mathcal{B}=0$.
With this restriction, the anomaly cancellation is fulfilled by imposing 
$\mathcal{A}_2=0$ as $\mathcal{A}_1 \propto \mathcal{B} = 0$.
As mentioned earlier, both $\mathcal{A}_2$ and $\mathcal{A}_1$
separately tend to zero in the classical limit, making this decomposition 
meaningful.  

In this case, the Poisson bracket between the hamiltonian and diffeomorphism
constraints is just $\left\{S^Q[N], D^Q[N^a]\right\}=0$. The conditions 
$\mathcal{B}=0$ and $\mathcal{A}_2=0$ can be translated into expressions 
for the $\alpha_1$ and $\alpha_2$ functions:
\begin{eqnarray}
\alpha_1 &=& -1+4(1-\beta) \frac{\mathbb{K}[2]\mathbb{K}[v_2]}{\mathbb{K}[1]^2}
-2\frac{\mathbb{K}[v_2]^2}{\mathbb{K}[1]^2}  
+ (4\beta-1)\cos(v_2 \bar{\mu} \gamma \bar{k}), \label{alpha1B0} \\
\alpha_2 &=&-1+\frac{2 \mathbb{K}[2]-\mathbb{K}[v_2]}{\mathbb{K}[v_1]}. \label{alpha2B0}
\end{eqnarray} 
The expressions for $\alpha_1$ and $\alpha_2$  are parametrized by the  
integers $v_1$ and $v_2$ only. However, the dependence upon $v_1$   
vanishes when $\alpha_2$ is used in the hamiltonian constraint.

The derived expressions for $\alpha_1$ and $\alpha_2$ do contain $\mathbb{K}[n]$ 
functions in the denominators. In principle,  $\alpha_1$ and $\alpha_2$
could therefore diverge for some values of $\bar{k}$. However, in the counter-terms $S_{C1}$  
and $S_{C2}$, $\alpha_1$ is multiplied by $\mathbb{K}[1]^2$ and $\alpha_2$ by $\mathbb{K}[v_1]$. 
The subsequent cancellation prevents any physical divergence from occurring.

\section{Introducing matter}

We have shown that the gravity sector of the vector perturbations with holonomy corrections 
can be made anomaly free. We will now extend this result by introducing scalar matter. The 
matter Hamiltonian does not depend on the Ashtekar connection and is therefore not subject 
to holonomy corrections.  Furthermore, for vector perturbations, $\delta N = 0$. The 
matter Hamiltonian is perturbed up to the second order as follows:
\begin{equation}
H_{\rm{m}}[N] = \bar{H}_{\rm m } + \delta H_{\rm m } = \int_{\Sigma} d^3x \bar{N} ( C_{\rm m}^{(0)}+C_{\rm m}^{(2)}), 
\end{equation}
where 
\begin{equation}
C_{\rm{m}}^{(0)} = \bar{p}^{3/2} \left[  \frac{1}{2} \frac{\bar{\pi}^2}{\bar{p}^3}+V(\bar{\varphi}) \right].
\end{equation} 
The value of $C_{\rm{m}}^{(2)}$ is given by 
\begin{eqnarray}
C_{\rm{m}}^{(2)}  &=& \frac{1}{2} \frac{\delta \pi^2}{\bar{p}^{3/2}}
+\frac{1}{2} \sqrt{\bar{p}}\delta^{ab} \partial_a \delta \varphi \partial_b \delta \varphi 
+\frac{1}{2}\bar{p}^{3/2} V_{,\varphi \varphi}(\bar{\varphi}) \delta\varphi^2  \nonumber \\
&+&\left(  \frac{1}{2} \frac{\bar{\pi}^2}{\bar{p}^{3/2}}-\bar{p}^{3/2} V(\bar{\varphi}) \right)
\frac{\delta^k_c \delta^j_d \delta E^c_j \delta E^d_k }{4\bar{p}^2},
\end{eqnarray}
where we have used the condition $\delta^i_a \delta E^a_i = 0$.
The matter diffeomorphism constraint is given by:
\begin{equation}
D_{\rm{m}}[N^a] =\int_{\Sigma} d^3x \delta N^a \bar{\pi}(\partial_a \delta \varphi).  
\end{equation}

The total hamiltonian and diffeomorphism constraints are
\begin{eqnarray}
S_{\rm{tot}}[N] &=& S^Q[N]+H_{\rm{m}}[N], \\
D_{\rm{tot}}[N^a] &=&D^Q[N^a] +D_{\rm{m}}[N^a]. 
\end{eqnarray}
The resulting Poisson brackets are the following:
\begin{eqnarray}
\left\{S_{\rm{tot}}[N_1], S_{\rm{tot}}[N_1] \right\} =  0, \\
\left\{D_{\rm{tot}}[N^a_1], D_{\rm{tot}}[N^a_2] \right\} = 0,  \\ 
\left\{S_{\rm{tot}}[N], D_{\rm{tot}}[N^a]\right\} = \frac{\bar{N}}{\sqrt{\bar{p}}} \mathcal{B} D^Q[N^a] 
+\frac{\bar{N}}{\kappa\sqrt{\bar{p}}} \int_{\Sigma}  d^3x\delta N^c \delta^k_c  
(\partial_d  \delta E^d_k) \delta E^d_k \mathcal{A} \nonumber \\
+[\cos(v_2 \bar{\mu} \gamma \bar{k} )-1]
\frac{\sqrt{\bar{p}}}{2}\left(\frac{\bar{\pi}^2}{2\bar{p}^3}-V(\bar{\varphi})\right)
\int_{\Sigma} d^3x \bar{N}   \partial_c (\delta N^a) \delta^j_a \delta E^c_j  \nonumber \\
+  \frac{\bar{\pi}}{\bar{p}^{3/2}} \int_{\Sigma}d^3x \bar{N} (\partial_a \delta N^a  )\delta \pi 
-\bar{p}^{3/2}V_{\varphi}(\bar{\varphi}) \int_{\Sigma}d^3x \bar{N} (\partial_a \delta N^a  ) \delta \varphi.
\end{eqnarray}
Anomaly freedom requires $\mathcal{B}=0$, $\mathcal{A}=0$, $v_2=0$ (classical diffeomorphism constraint), 
and also $\delta \varphi = 0 = \delta \pi$. The latter conditions  $\delta \varphi = 0 = \delta \pi$ are due to the fact that metric scalar perturbations are not considered. 
Consistently, scalar field perturbations are vanishing too.  In fact, one could set 
$\delta \varphi = 0 = \delta \pi$ from the very beginning but, without assuming this, it can 
be shown that the condition $\delta \varphi = 0 = \delta \pi$ in fact resulting from the 
anomaly freedom.

The associated counter-terms are  given by (\ref{alpha1B0}) and (\ref{alpha2B0})
with $v_2=0$. Two non-vanishing counter-terms are required in contrast to the  
gravity sector, where only one  counter-term was sufficient to fulfill the anomaly freedom conditions. 
The integer $v_1$ remains undetermined but the dependence upon this parameter 
cancels out in the hamiltonian constraint. Namely, applying the counter-terms  
(\ref{alpha1B0}) and (\ref{alpha2B0}) with $v_2=0$, we find that the anomaly free 
hamiltonian constraint is given by:
\begin{equation}
S^Q_{\rm{free}}[N] = \frac{1}{2 \kappa} \int_{\Sigma} d^3x 
\left[ \bar{N}(C^{(0)}_{\rm{free}} +C^{(2)}_{\rm{free}} )\right],  \label{scalarfree}
\end{equation}
where 
\begin{eqnarray}
C^{(0)}_{\rm{free}}  &=& -6\sqrt{\bar{p}}  \left(\mathbb{K}[1] \right)^2,  \\
C^{(2)}_{\rm{free}}  &=&  -\frac{1}{2\bar{p}^{3/2}} \left[ 4(1-\beta)\mathbb{K}[2]\bar{k} -2\bar{k}^2 +(4\beta-1)\mathbb{K}[1]^2 \right]
(\delta E^c_j \delta E^d_k \delta^k_c \delta^j_d)  \nonumber \\
&+&  \sqrt{\bar{p}}(\delta K^j_c \delta K^k_d \delta^c_k \delta^d_j) 
-\frac{2}{\sqrt{\bar{p}}}\left(2\mathbb{K}[2] -\bar{k}\right)(\delta E^c_j \delta K^j_c ). 
\label{anomalyfreeC2}
\end{eqnarray}
The gravitational diffeomorphism constraint holds its classical 
form($v_2=0$). This is in agreement with LQG expectations.
Interestingly, this can also be obtained here as a result of  anomaly freedom.

The obtained anomaly-free Hamiltonian (\ref{scalarfree}) is determined 
up to the choice of the $\bar{\mu}$ functions. There are no other remaining 
ambiguities. The $\bar{\mu}$ function appears in definition of the $\mathbb{K}[n]$ 
function. Because of this, there is also explicit appearance of the factor $\beta$ 
in equation (\ref{anomalyfreeC2}). The choice $\beta =-1/2$ is preferred by various 
considerations \cite{Nelson:2007um}.
Recently, this value was shown to be required also by the conditions on the anomaly-free 
scalar perturbations with holonomy corrections \cite{Scalars2011}.  For this choice
of the $\beta$ parameter, the remaining freedom is a parameter of proportionality in
relation $\bar{\mu} \propto \bar{p}^{-1/2}$. This parameter can be written 
as $\sqrt{\Delta}$, so $\bar{\mu} = \sqrt{\Delta/\bar{p}}$. The parameter $\Delta$
has interpretation of physical area, around which the elementary holonomy 
is defined. It is expected that $\Delta \sim l_{Pl}^2$, where $l_{Pl}$ is the 
Planck length. However,  determination of the accurate value of $\Delta$ 
is a subject to empirical verifications. 
 
It is worth noticing about the Hamiltonian constraint (\ref{scalarfree})
that the effective holonomy corrections, due to the counter-terms, are no 
longer \emph{almost periodic functions},  defined as follows \cite{Ashtekar:2003hd}
\begin{equation}
f(\bar{k}) = \sum_{n}\xi_n e^{i \bar{\mu} \gamma \bar{k} n}.
\end{equation}
In this expression, $n$ runs over a finite number of integers and $\xi_n\in \mathbb{C}$.  
This does not lead  to any problem at the classical level. However, 
difficulties may appear when going to the quantum theory on  lattice states. 
This is because the quantum operator corresponding to $\bar{k}$ does not 
exist in contrast to  the $\mathbb{K}[n]$ functions, which are  almost periodic 
functions.  This problem does not exist if the gravitational sector, without any 
matter content, is considered alone. However, the diffeomorphism constraint then has
to be holonomy corrected, as studied previously. In such a case, the background 
terms in the anomaly-free gravitational Hamiltonian are  almost periodic functions.
The loop quantization can therefore be directly performed.  

\section{Gauge invariant variable}

The coordinate transformation $x^{\mu} \rightarrow x^{\mu} +\xi^{\mu}$ generates a tensor gauge transformation. In the case of vector modes, the coordinate transformation is parametrized by the shift vector  
$N^a = \xi^a$, where ${\xi^a}_{,a}=0$, Therefore, the resulting gauge transformation is generated 
by the diffeomorphism constraint $\delta_{\xi} f= \{ f, D^Q[\xi^a] \}$. The corresponding transformations for the canonical variables 
are:
\begin{eqnarray}
\delta_{\xi} (\delta E^a_i) &=& \{\delta E^a_i, D^Q[\xi^a] \}=- \bar{p} \partial_i \xi^a  \label{xiE},   \\
\delta_{\xi} (\delta K^i_a) &=& \{\delta K^i_a, D^Q[\xi^a] \}= \mathbb{K}[v_2] \partial_a \xi^i \label{xiK}.
\end{eqnarray}
Based on the equation of motion  $\dot{E}^a_i= \{E^a_i, H_{\rm{G}} \}$, and the definition (\ref{deltaE}),
one finds the expression of $\delta K^i_a$.  The dot means differentiation with respect to the conformal 
time since we have chosen $\bar{N}=\sqrt{\bar{p}}$. Using equations (\ref{xiE}) and (\ref{xiK}) one finds:
\begin{eqnarray}
\delta_{\xi} F^a &=&  \xi^a, \\
\delta_{\xi} S^a &=&  \dot{\xi}^a +(2 \mathbb{K}[2] -\mathbb{K}[v_1](1+\alpha_2)-\mathbb{K}[v_2]) \xi^a.
\end{eqnarray}
Based on this, one can define a gauge invariant variable
\begin{equation}
\sigma^a := S^a- \dot{F}^a-
\underbrace{(2 \mathbb{K}[2] -\mathbb{K}[v_1](1+\alpha_2)-\mathbb{K}[v_2])}_{=-\mathcal{B}} F^a,
\end{equation}
such that $\delta_{\xi}\sigma^a = 0$.

\section{Equations of motion}

In this section we derive the equation of motion for the gauge-invariant variable found in the 
previous section. 

For the sake of completeness, we recall that the equations of motion for the background part are:
\begin{eqnarray}
\dot{\bar{p}} &=&  \bar{N}2\sqrt{\bar{p}} (\mathbb{K}[2]),    \\
\dot{\bar{k}}  &=&  - \frac{\bar{N}}{\sqrt{\bar{p}}}\left[ \frac{1}{2} (\mathbb{K}[1])^2+\bar{p}\frac{\partial}{\partial \bar{p}}  (\mathbb{K}[1])^2 \right] 
\nonumber \\
&+& \frac{\kappa}{3V_0} \left(\frac{\partial \bar{H}_{\rm{m}}}{\partial \bar{p}} \right),  
\end{eqnarray}
where $\bar{H}_{\rm{m}} = V_0 \bar{N} C_{\rm{m}}^{(0)} $ and $\bar{N}=\sqrt{\bar{p}}$.  
For a free scalar field, an analytical solution to these equations can be found
\cite{Mielczarek:2008zv}: 
\begin{equation}
\bar{p} = \left(\frac{1}{6} \gamma^2 \Delta \pi_{\varphi}^2 \kappa + \frac{3}{2} \kappa \pi_{\varphi}^2 t^2 \right)^{1/3}.
\end{equation}
This solution represents a symmetric bounce.

The diffeomorphism constraint $\frac{\delta }{\delta \delta N^a}  D_{\rm{tot}}[N^a]= 0$  leads to
the equation
\begin{equation}  
\bar{p} (\partial_k \delta K^k_a ) + \left( \mathbb{K}[v_2]\right)\delta^k_a  (\partial_d \delta E^d_k )
= \kappa \bar{\pi} \partial_a(\delta \varphi).
\label{diffconstr}
\end{equation}
Using the symmetrized variables 
\begin{eqnarray}
\delta {K^{(i}}_{a)} &=& \frac{1}{2}\left[  (2 \mathbb{K}[2]-\mathbb{K}[v_1](1+\alpha_2))\left( {F_{a,}}^i+{F^i}_{,a}\right) \right. \nonumber  \\
&+& \left. \left( {F_{a,}}^i+{F^i}_{,a} \right)\dot{}-\left( {S_{a,}}^i+{S^i}_{,a} \right) \right]  \nonumber \\
&=& -\frac{1}{2} \left(  {\sigma_{a,}}^i+{\sigma^i}_{,a}\right)+\frac{1}{2} \mathbb{K}[v_2]\left( {F_{a,}}^i+{F^i}_{,a}\right), 
\label{deltaK}
\end{eqnarray} 
and 
\begin{equation}
\delta {E^{(i}}_{a)} = -\bar{p} \frac{1}{2}\left( {F_{a,}}^i+{F^i}_{,a}\right),  
\end{equation}
equation (\ref{diffconstr}) can be rewritten as
\begin{equation}
- \frac{\bar{p}}{2}\nabla^2 \sigma_a = \kappa \bar{\pi} \partial_a(\delta \varphi). \label{nablasigma}
\end{equation}
Because $\delta \varphi =0$ (from the anomaly-free condition), the symmetric 
diffeomorphism constraint simplifies to  the Laplace equation $\nabla^2 \sigma_a =0$.
Since, the spatial slice is flat ($\Sigma = \mathbb{R}^3$)
there are no boundary conditions on $\sigma_a$. This restricts the possible solutions
of the Laplace equation to $\sigma_a = b_a +d_a^{c}x_c$, where $b_a$ and $d_a^{c}$ 
are sets of constants.  However, because $\sigma_a$ is a perturbation (there is no contribution
from the zero mode), 
\begin{equation}
\int_{\Sigma} d^3x \sigma_a = 0, \label{condition}
\end{equation} 
as required from the consistency of the perturbative expansion. This is also the reason 
why the first order perturbation of the Hamiltonian is vanishing, $\int_{\Sigma}  C^{(1)} d^3 x=0$. 
Condition (\ref{condition}) implies   $b_a=0$ and $d_a^{c}=0$, which leads to 
$\sigma_a=0$. This shows that our gauge invariant variable $\sigma_a$ is 
identically equal to zero in absence of  vector matter, in agreement with earlier studies 
\cite{Langlois:1994ec}. This can also be proved by expanding $\sigma_a$  into  Fourier 
modes. 

In order to have non-vanishing (physical) vector modes  $\sigma_a$, a source 
term in equation (\ref{nablasigma}) therefore has to be present.  
With "vector matter", this reads as  \cite{Bojowald:2007hv}:
\begin{equation}
-\frac{1}{2\bar{p}} \nabla^2 \sigma_a =  8\pi G (\rho+P)V_a, 
\end{equation}
where $\rho$ and $P$ are the energy density and pressure of the vector matter and 
$V_a$ is a matter perturbation vector. If $(\rho+P)V_a \neq 0$ then
$\sigma_a\neq 0$ so  physical vector perturbations are expected. However, it should be pointed
out that proving that the formulation remains anomaly-free
in presence of the vector matter remains an open issue. 
This could be checked,{\it e.g.}, by introducing an electromagnetic field in the Hamiltonian 
formulation \cite{Bojowald:2007gt}. We leave this problem to be analyzed elsewhere.

Due to the Gauss constraint, we introduce the symmetrized  variable
\begin{equation}
\mathfrak{S}^i_a := {\sigma^i}_{,a}+{\sigma_{a,}}^i.
\end{equation}
The equation of motion for this variable reads as:
\begin{equation}
-\frac{1}{2} \frac{d}{d\eta}\mathfrak{S}^i_a-\frac{1}{2}( 2\mathbb{K}[2]+\mathcal{B} )\mathfrak{S}^i_a 
+\mathcal{A} {F^{(i}}_{ ,a)}=\kappa \bar{p} \delta T^{(i}_{a)},
\label{vecteqfull}
\end{equation}
where 
\begin{equation}
\delta T^{i}_{a} =
\frac{1}{\bar{p}}\left[\left( \frac{1}{3V_0} \frac{\partial \bar{H}_m}{\partial \bar{p}}   \right) 
\left(  \frac{\delta E^c_j \delta^j_a \delta^i_c}{ \bar{p}} \right)  +\frac{\delta H_{m}}{\delta \delta E^a_i}  \right].
\end{equation}
For scalar matter $\delta T^{i}_{a}=0$. The same holds for tensor modes 
\cite{Mielczarek:2009vi} (the reasons are the same because 
$\delta^i_a \delta E^a_i =0$ and $\delta N =0$). When imposing the anomaly freedom 
conditions $\mathcal{A}=0$ and $\mathcal{B}=0$, equation (\ref{vecteqfull}) simplifies to
\begin{equation}
-\frac{1}{2} \frac{d}{d\eta}\mathfrak{S}^i_a-\frac{1}{2}\underbrace{( 2\mathbb{K}[2])}_{= \frac{1}{\bar{p}} \frac{d \bar{p}}{d\eta}  }
\mathfrak{S}^i_a = 0,
\label{vectoreq}
\end{equation}
with fully determined coefficients. Of course without vector mater, as discussed above, the 
variable $\mathfrak{S}^i_a$ is equal to zero and the equation (\ref{vectoreq}) is trivially 
satisfied. However, the
presence on a non-vanishing contribution from $V_a$ allows for  non-trivial solutions of
 equation 
(\ref{vectoreq}). In such a case, equation (\ref{vectoreq}) leads to:
\begin{equation}
\mathfrak{S}^i_a = \frac{\rm const }{\bar{p}} = \frac{\rm const }{a^2}.
\end{equation}
For a symmetric bounce driven by a free scalar field:
\begin{equation}
\mathfrak{S}^i_a \propto \frac{1}{\left(\frac{2\pi}{3\sqrt{3}} \gamma^3 l^2_{\rm Pl} + t^2 \right)^{1/3}}.
\end{equation}
The evolution is smooth through the bounce. The amplitude of the perturbations grows 
during the contraction and decreases in the expanding phase. The maximum amplitude 
is reached at the transition point (bounce). Moreover, this evolution is independent on the 
length of the considered mode, as can be seen by performing a Fourier transform of 
the function $\sigma_a$. Because of this, there is significant difference with respect to 
tensor and scalar perturbations. For the scalar and tensor perturbations, the evolution is 
different depending on whether the mode length is shorter or longer that the Hubble horizon. 
In particular, on  super-horizon scales, the amplitude of the scalar and tensor perturbations 
is \emph{frozen}. In contrast, for the vector modes there is no such effect.  Therefore, in an
expanding universe, the amplitude of vector modes decreases with respect to the super-horizon 
tensor and scalar perturbations. The contribution  from  vector modes becomes 
negligible during the expansion phase. However, the situation reverses  in the contracting 
phase, before the bounce.  Then, the amplitude of the vector perturbations grows with 
respect to the super-horizon tensor and scalar perturbations. Therefore, on  very large 
scales the vector perturbations can play an important role, {\it e.g.} leading to the generation of 
large scale magnetic fields \cite{Battefeld:2004cd}. This could lead to a new tool to explore 
physics of the (very) early universe.

\section{Summary and conclusions}

In this paper we have studied the issue of anomaly cancellation for 
vector modes with holonomy corrections in LQC. Our strategy is based 
on the introduction of counter-terms in the holonomy-corrected 
hamiltonian constraint. In our study, we have also introduced possible holonomy 
corrections to the diffeomorphism constraint. We have shown, first, 
that the anomaly cancellation cannot be achieved without 
counter-terms. Holonomy corrections to the 
diffeomorphism constraint do not help significantly to fulfill the
anomaly freedom conditions, that are anyway satisfied up to the fourth order 
in the canonical variable $\bar{k}$. Then, we have studied 
the anomaly issue for the gravitational sector with two counter 
terms. We have shown that the conditions of anomaly freedom can 
be met with at least one non-vanishing counter-term. The resulting effective holonomy  
corrections are almost periodic functions 
only if the  diffeomorphism constraint  is holonomy corrected.  
Subsequently, we have investigated the issue of anomaly cancellation 
when a matter scalar field is added. In this case, the closure
conditions are more restrictive and fully determine the form of the 
resulting hamiltonian constraint. Moreover, this requires 
that the  diffeomorphism constraint holds its classical form, in agreement with LQG expectations. 
Because of this, the  effective holonomy corrections, which take 
into account contributions from the counter-terms, are no more almost 
periodic functions.  
We have found the gauge invariant variable and the corresponding 
equation of motion. The solution to this equation were also given. 
We have analyzed this solution for the symmetric bounce model to point out that
the vector perturbations smoothly pass through the bounce, 
where their amplitude reaches its maximum but finite value.
The work performed here for scalar matter should be extended to vector 
matter to fully address the considered issue.

In \cite{Scalars2011}, we address the related issue of anomaly 
freedom for scalar perturbations with holonomy corrections. This is 
most important from the observational viewpoint.

\ack

TC and JM were supported from the Astrophysics Poland-France (Astro-PF).
JM has been supported by Polish Ministry of Science and Higher Education 
grant N N203 386437 and by Foundation of Polish Science award START.

\end{document}